\documentclass[12pt,english,floatfix,showkeys,superscriptaddress,aps,prd,preprint]{revtex4}
\usepackage[T1]{fontenc}
\usepackage{lmodern}
\usepackage{amsmath}
\usepackage{amssymb}
\usepackage{graphicx}
\usepackage{float}
\usepackage{esint}
\usepackage{subcaption}
\usepackage{dcolumn}
\usepackage{babel}
\usepackage{csquotes}
\usepackage{color}
\usepackage{slashed}
\usepackage{simplewick}
\usepackage{amsmath,latexsym}

\usepackage{hyperref}
\hypersetup{
    colorlinks,
    citecolor=blue,
    filecolor=green,
    linkcolor=purple,
    urlcolor=red,
}

\usepackage{slashed}

\usepackage{hyperref}
\hypersetup{colorlinks,breaklinks,
			citecolor=[rgb]{0,0.0,1.0},
            urlcolor=[rgb]{0.0,0.0,1.0},
            linkcolor=[rgb]{0,0.5,0.9}}

\begin{document}

\title{Casimir effect in a Lorentz-violating tensor extension of a scalar field theory}

\author{M. C. Ara\'{u}jo}
\email{michelangelo@fisica.ufc.br}
\affiliation{Universidade Federal do Cear\'a (UFC), Departamento de F\'isica,\\ Campus do Pici, Fortaleza - CE, C.P. 6030, 60455-760 - Brazil.}
\author{J. Furtado}
\email{job.furtado@ufca.edu.br}
\affiliation{Universidade Federal do Cariri (UFCA), Av. Tenente Raimundo Rocha, \\ Cidade Universit\'{a}ria, Juazeiro do Norte, Cear\'{a}, CEP 63048-080, Brasil}
\author{R. V. Maluf}
\email{r.v.maluf@fisica.ufc.br}
\affiliation{Universidade Federal do Cear\'a (UFC), Departamento de F\'isica,\\ Campus do Pici, Fortaleza - CE, C.P. 6030, 60455-760 - Brazil.}
\affiliation{Departamento de F\'{i}sica Te\'{o}rica and IFIC, Centro Mixto Universidad de Valencia - CSIC. Universidad de Valencia, Burjassot-46100, Valencia, Spain.}



\date{\today}

\begin{abstract}
This paper investigates the Casimir Energy modifications due to the Lorentz-violating CPT-even contribution in an extension of the scalar QED. We have considered the complex scalar field satisfying Dirichlet boundary conditions between two parallel plates separated by a small distance. An appropriate tensor parametrization allowed us to study the Casimir effect in three setups: isotropic, anisotropic parity-odd, and anisotropic parity-even. We have shown that the Lorentz-violating contributions promote increased Casimir energy for both the isotropic and anisotropic parity-odd configurations. However, in the parity-even case, the Lorentz-violating terms can promote either an increase or a decrease in the Casimir energy. We have shown that both the increased and decreased amounts in the Casimir energy depend on the momentum projection over the Lorentz-violating vectors.
\end{abstract}

\keywords{Casimir effect, Lorentz symmetry breaking, CPT violation, Lorentz-violating scalar QED.}

\maketitle

\section{Introduction}\label{intro}

The Casimir effect is a purely quantum phenomenon and presents as one of the most direct manifestations of the existence of vacuum quantum fluctuations \cite{Bordag:2001qi,Bordag:2009zz}. Predicted by H. Casimir in 1948 \cite{Casimir:1948dh}, the Casimir effect is characterized by the force of attraction taking place between two parallel, electrically neutral conducting plates separated by a short distance in the quantum vacuum, and it emerges as a consequence of the boundary conditions imposed on the electromagnetic field by the plates. These boundaries force the frequencies of the field to be in a discrete spectrum where only specific, well-defining values are allowed. Consequently, the zero-point energy (vacuum energy) also goes through changes. After some regularization procedure, we obtain a finite amount that can be interpreted as the needed energy to keep the plates in the desired configuration. Experimentally, M. J. Sparnaay confirmed the Casimir effect at the micrometer scale by Al plates in 1958 \cite{Sparnaay:1958wg}. About thirty-nine years later, the experiment was again carried out with higher accuracy  using layers of Cu and Au by Lamoreaux \cite{Lamoreaux:1996wh} and for a metallic sphere and flat plate by Mohideen and Roy \cite{Mohideen:1998iz}. A modern review of the experimental methods and realistic measurements can be found in the references \cite{Mohideen:1998iz,Klimchitskaya:2009cw,Milton:2019izg}.   

Although the Casimir effect was initially studied for the electromagnetic field, it is known to occur for any quantum field under certain boundary conditions. These boundaries can be material means, interfaces between two phases of the vacuum, or even space-time topologies \cite{Bordag:2001qi,Plunien:1986ca}. In fact, the Casimir force can depend on many parameters: as the geometry of the confining materials and the type of boundary conditions (Dirichlet, Neumann, or
mixed) \cite{Cruz:2017kfo,Muniz:2015jba}, or even the presence of extra-dimensions \cite{Ponton:2001hq}. In this sense, the Casimir effect was widely evaluated in a number of different scenarios: as black holes  \cite{Muniz:2015jba,Alencar:2018avb,Sorge:2019ecb,Christodoulakis:2001ps,Vagenas:2003tv}, modified gravity \cite{Lambiase:2016bjy,Buoninfante:2018bkc}, Ho\v{r}ava-Lifshitz-like field theory \cite{Muniz:2013uva,Ferrari:2010dj,MoralesUlion:2015tve,Muniz:2014dga,daSilva:2019iwn}, and Lorentz-violation contexts \cite{Cruz:2017kfo,Blasone:2018nfy,Escobar-Ruiz:2021dxi,Cruz:2018thz}. 

As is known, Lorentz covariance and CPT symmetry are preserved by the Standard Model (SM) of particle physics. However, although experimentally successful, the SM still leaves some open questions, such as the neutrino oscillation \cite{Barger:2012pxa}, or the lack of a quantum description of gravity, making the search for a physics beyond the SM a quite natural choice. In these new scenarios, Lorentz and CPT violation could arise from an underlying theory combining gravity with quantum mechanics such as string theory \cite{Kostelecky:1988zi,Kostelecky:1991ak},  Horava-Lifshitz gravity \cite{Horava:2009uw,Cognola:2016gjy}, loop quantum gravity \cite{Gambini:1998it,Bojowald:2004bb}, among others. Alternatively, in order to investigate signs of such violations at low energies, Colladay and Kosteleck\'{y} proposed the Standard Model Extension (SME) \cite{Kostelecky:1994rn,Colladay:1996iz,Colladay:1998fq,Kostelecky:2003fs}. As a comprehensive framework for treating CPT violation at the level of effective field theory, the SME has led to several experimental investigations \cite{Link:2002fg,BaBar:2007run,KLOE-2:2013ozx,KLOE:2010yad,D0:2015ycz,LHCb:2016vdl}. Moreover, it has
been extensively studied in numerous contexts such as neutrino oscillation \cite{MINOS:2008fnv,MINOS:2010kat,Katori:2012pe,MiniBooNE:2011pix}, radiative corrections \cite{Jackiw:1999yp,Kostelecky:2001jc, Ferrari:2021eam, Furtado:2014cja}, supersymmetric models \cite{Lehum:2018tpi,Ferrari:2017rwk,Belich:2015qxa,Colladay:2010tx}, and noncommutative quantum field theories \cite{Carroll:2001ws,Hayakawa:1999yt}.

More recently, Edwards and Kosteleck\'{y} proposed an extension of the SM scalar sector where small CPT deviations in neutral-meson oscillations could be evaluated \cite{Edwards:2019lfb}. Structurally, the extension consists of a general effective scalar field theory in any spacetime dimension that contains explicit-perturbative spin-independent Lorentz-violating operators of arbitrary mass dimension \cite{Edwards:2018lsn}. The significance of spin-independency lies on the  fact that even for a particle with nonzero spin, Lorentz-violating effects could be handled as though the particle had zero spin. This is extremely interesting since the Higgs boson is the only example of a fundamental spinless particle in the SM.

In this paper, the four-dimensional tensor sector of the Lorentz violating extension of the scalar electrodynamics proposed in Ref. \cite{Edwards:2018lsn} will be considered in the calculation of the Casimir energy for a massive complex scalar field satisfying Dirichlet
boundary conditions between  two large parallel plates separated by a small distance. As will be shown, by an appropriate  tensor parametrization in terms of vectors comprising the Lorentz-violating coefficients, we will be able to analyze the Casimir effect in three different scenarios: isotropic, anisotropic parity-odd, and anisotropic parity-even scenarios. Consequently, effects of Lorentz violation on the Casimir energy may be direction-dependent. In all the cases, we apply the dimensional regularization process to regularize the Casimir energy.     

This paper is organized as follows. In Sec. \ref{secdois}, the model is presented and the tensor parameterization is discussed. In Sec. \ref{sec3}, the isotropic case is considered. Sections \ref{sec4} and \ref{sec5} are respectively dedicated to the anisotropic parity-even and anisotropic parity-odd sectors. In Sec. \ref{conclusion}, the main results of the paper are summarized.

\section{The model}\label{secdois}

In this section, we present some features concerning the model we are going to investigate the Casimir effect. As proposed by Edwards and Kosteleck\'{y} in reference \cite{Edwards:2018lsn}, the Lagrange density describing the propagation of a complex scalar field $\phi$ of mass $m$ in the presence of arbitrary Lorentz-violating effects can be written in the form

\begin{eqnarray}\label{densidadelagmodelocomplescalar}
\mathcal{L} = \partial_{\mu}\phi^{\ast}\partial^{\mu}\phi + \hat{K}_c^{\mu \nu}\partial_{\mu}\phi^{\ast}\partial_{\nu}\phi - \frac{i}{2}\left[ \hat{K}^{\mu}_a (\phi^{\ast}\partial_{\mu}\phi - \partial_{\mu}\phi^{\ast}\phi ) \right] - m^2\phi^{\ast}\phi,
\end{eqnarray} where $\hat{K}_c^{\mu \nu}$ and $\hat{K}^{\mu}_a$ are respectively a null trace constant tensor and a constant vector assumed to introduce only small Lorentz violation to conventional physics. In this work, our focus will be on the tensor sector of the model \eqref{densidadelagmodelocomplescalar}. General vector contributions will be left as perspective. Moreover, Lorentz violation effects on the Casimir energy due to similar vector terms in a massive real scalar field model were investigated in Ref. \cite{Cruz:2017kfo}.

The corresponding modified Klein-Gordon equation is
\begin{eqnarray}\label{modifkleing}
( \Box + m^2 + \hat{K}_c^{\mu \nu}\partial_{\mu}\partial_{\nu})\phi = 0,
\end{eqnarray} with similar equation for $\phi^{\ast}$.

\begin{figure}[!ht]
    \centering
    \includegraphics[scale=0.4]{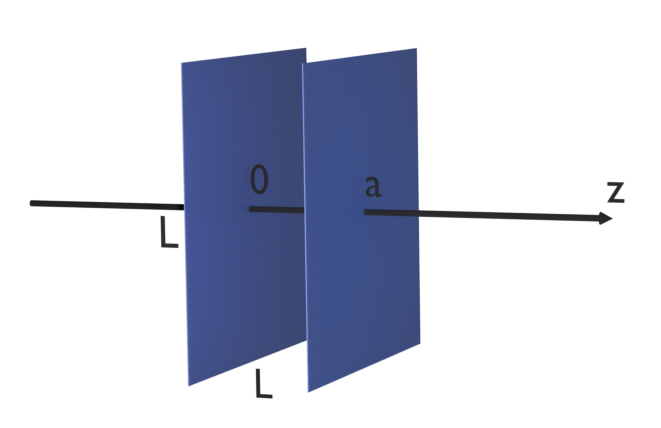} 
        \caption{Two parallel plates with area $L^2$ separated by a small distance $a$. We mean here by ``small'' that $a\ll L$.}
    \label{fig1}
\end{figure}

Now, since $\hat{K}_c^{\mu \nu}$ is a symmetric traceless tensor, we can use the general parametrization
\begin{eqnarray}\label{parametrizacaokemuv}
\hat{K}_c^{\mu \nu} = \frac{1}{2}(U^{\mu}V^{\nu} + U^{\nu}V^{\mu}) - \frac{1}{4}\eta^{\mu \nu}(U\cdot V),
\end{eqnarray} where $U$ and $V$ are two arbitrary 4-vectors which comprise the Lorentz-violating coefficients \cite{Casana:2010nd}. This parametrization is useful to investigate the different configurations of this tensor individually \cite{dosSantos:2018rns}: the anisotropic parity-even sector, for example, is
parameterized by two pure spacelike 4-vector $U = (0,\mathbf{u})$ and $V = (0,\mathbf{v})$, with $\mathbf{u}\cdot \mathbf{v} = 0$, whereas the
anisotropic parity-odd sector is represented by $U = (0,\mathbf{u})$ and $V = (v_0,\mathbf{0})$. The isotropic sector is in turn recovered by two timelike 4-vector $U = (u_0,\mathbf{0})$ and $V = (v_0,\mathbf{0})$.
In the following sections, we will calculate the vacuum energy for each sector above considering  the complex scalar field satisfying Dirichlet boundary conditions,
\begin{eqnarray}\label{conditionsdirichlet}
\phi(x,y,0,t) = \phi(x,y,a,t) = 0,
\end{eqnarray}
 between two large parallel plates separated by a small distance $a$, see Fig. \eqref{fig1}.

\section{The isotropic sector}\label{sec3}

As discussed before, the tensor $\hat{K}^{\mu \nu}_c$ in this case must be parameterized by two timelike 4-vectors $U = (u_0,\mathbf{0})$ and $V = (v_0,\mathbf{0})$ so that Eq. \eqref{modifkleing} can be rearranged as
\begin{eqnarray}
(\Box + m^2 + \frac{3}{4}\, u_0\, v_0\, \partial_0^2 + \frac{1}{4}\, u_0\, v_0\, \nabla ^2)\phi = 0.
\end{eqnarray} 

Adopting the standard procedure \cite{Ryder:1985wq}, the quantum field can be expanded as
\begin{eqnarray}\label{solcampoisotrosector}
\phi(x) = \sqrt{\frac{2}{a}} \sum_{n=1}^{\infty}\int \frac{d^2p}{(2\pi)^2}\frac{\sin{(n\pi z/a)} }{2\omega_n(\mathbf{p})} \left[ a_n({\mathbf{p}})e^{-ip\cdot x} + b^{\dagger}_n({\mathbf{p}})e^{ip\cdot x}  \right],
\end{eqnarray} where $n$ is an integer and the angular frequency written as
\begin{eqnarray}\label{omegapnisotropico}
\omega_{n}(\mathbf{p}) = \frac{1}{\sqrt{1+\frac{3}{4}u_0 v_0}}\sqrt{\left( 1-\frac{1}{4}u_0 v_0 \right)\left[ \mathbf{p}^2 + \left( \frac{n\pi}{a} \right)^2 \right] + m^2}, 
\end{eqnarray} with $p\cdot x \equiv \omega_{n}(\mathbf{p})t - \mathbf{p}\cdot \mathbf{x}$, and $\mathbf{p} = (p_x,p_y)$. Note that the conditions \eqref{conditionsdirichlet} are directly verified.

Now, as a result of canonical quantization,  the Hamiltonian can be settled as \begin{eqnarray}\label{hamiloniana isotropico123}
H = \sum_{n=1}^{\infty} \int \frac{d^2p}{(2\pi)^2}\frac{1}{2}\left( 1+\frac{3}{4}u_0v_0 \right)\left[ a^{\dagger}_n({\mathbf{p}})a_n({\mathbf{p}}) + b^{\dagger}_n({\mathbf{p}})b_n({\mathbf{p}}) + \frac{L^2}{1+\frac{3}{4}u_0v_0} 2\omega_{n}(\mathbf{p})\right],
\end{eqnarray} where $L^2$ is the area of each plate and the annihilation and creation operators $a_n({\mathbf{p}})$ and $a^{\dagger}_n({\mathbf{p}})$, respectively, satisfy the algebra
\begin{eqnarray}
[a_n(\mathbf{p}),a^{\dagger}_m(\mathbf{q})] = \frac{1}{1+\frac{3}{4}u_0v_0}\, (2\pi)^2\, 2\omega_{n}(\mathbf{p})\, \delta^2(\mathbf{p} - \mathbf{q})\, \delta_{n m},
\end{eqnarray} with all the other commutation relations involving them being zero. Operators $b_n({\mathbf{p}})$ and $b^{\dagger}_n({\mathbf{p}})$ obey a similar algebra. Therefore, taking the vacuum expectation value of the Hamiltonian \eqref{hamiloniana isotropico123}, we find 
\begin{eqnarray}\label{e0iso}
E_{0\, (iso)} = \frac{L^2}{(2\pi)^2}\sum_{n=1}^{\infty}\int d^2p\,  \frac{1}{\sqrt{1+\frac{3}{4}u_0v_0}}\sqrt{\left(1-\frac{1}{4}u_0v_0\right)\left[ \mathbf{p}^2+\left(\frac{n\pi}{a}\right)^2 \right]+m^2},
\end{eqnarray} as the vacuum energy for the massive
complex scalar field in the isotropic sector. 

In order to reach a finite quantity of the equation \eqref{e0iso}, we will invoke the dimensional regularization procedure \cite{milton2001casimir} and write
\begin{eqnarray}\label{e0isoreg}
E^{Reg}_{0\, (iso)} = \frac{L^d}{(2\pi)^d}\sum_{n=1}^{\infty}\int d^dp\,  \frac{1}{\sqrt{1+\frac{3}{4}u_0v_0}}\sqrt{\left(1-\frac{1}{4}u_0v_0\right)\left[ \mathbf{p}^2+\left(\frac{n\pi}{a}\right)^2 \right]+m^2},
\end{eqnarray} where $d$ is the transverse dimension assumed as a continuous, complex variable. In this way we can perform the momentum integral by using the Schwinger proper time representation:
\begin{eqnarray}
\frac{1}{a^z}=\frac{1}{\Gamma(z)}\int_0^{\infty}dt\, t^{z-1}e^{-at}.
\end{eqnarray} 

\begin{figure}[h!]
    \centering
    \includegraphics{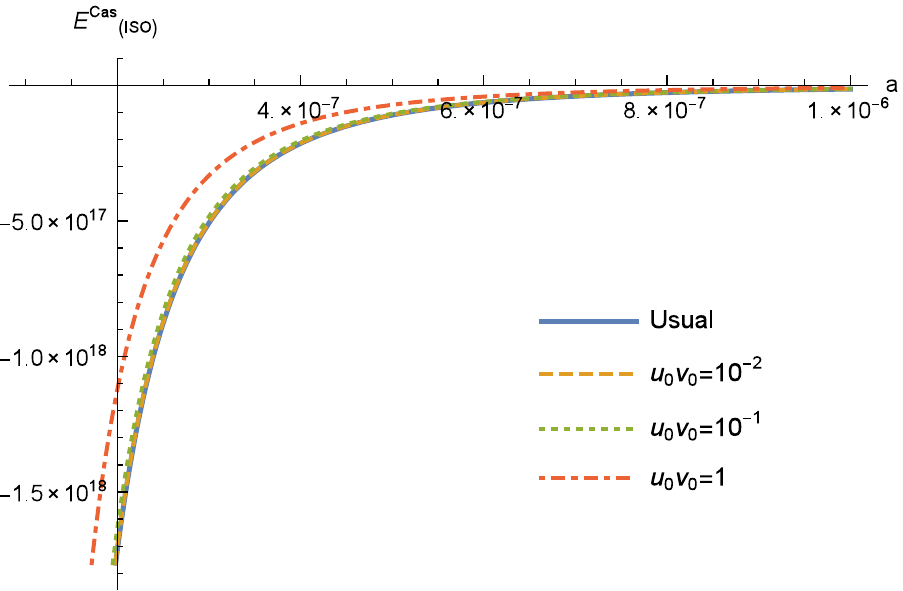}
    \caption{Casimir energy for some values of the product $u_0v_0$. For this plot we have considered $L=1$, $m=1$ and $a=(10^{-7},10^{-6})$.}
    \label{fig2}
\end{figure}

After performing the moment integration, the energy density becomes
\begin{eqnarray}\label{e0isoreg13}
E_{0\, (iso)}^{Reg} = \frac{L^d}{(2\pi)^d\, \sqrt{1+\frac{3}{4}u_0v_0}}\frac{\pi^{d/2}\, \Gamma \left( -\frac{d+1}{2} \right)}{\Gamma \left( -\frac{1}{2} \right)\left( 1-\frac{1}{4}u_0v_0 \right)^{d/2}}\sum_{n=1}^{\infty}\left[ m^2 + \left( 1-\frac{1}{4}u_0v_0 \right)\left( \frac{n\pi}{a} \right)^2 \right]^{\frac{d+1}{2}}.\nonumber\\
\end{eqnarray} One put the sum over $n$ into a more helpful form using an Epstein-Hurwitz Zeta function type \cite{Zhai:2010mr}  
\begin{eqnarray}
\sum_{n=-\infty}^{+\infty}(b n^2+\mu^2)^{-s} = \sqrt{\frac{\pi}{b}}\, \frac{\Gamma \left(s-\frac{1}{2}\right)}{\Gamma (s)}\mu^{1-2s}+\frac{2\pi^s}{\sqrt{b}\Gamma (s)}\sum_{n=-\infty}^{+\infty \, \prime}\mu^{\frac{1}{2}-s}\left( \frac{n}{\sqrt{b}} \right)^{s-\frac{1}{2}}K_{\frac{1}{2}-s}\left( \frac{2\pi \mu n}{\sqrt{b}} \right), \nonumber\\
\end{eqnarray} 
where $K_{\nu}(z)$ is the modified Bessel function, and the prime in the sum means that the term $n = 0$ has to be excluded. After some algebra, we can write the dimensionally
regularized energy in Eq. \eqref{e0isoreg13} as
\begin{eqnarray}
E_{0\, (iso)}^{Reg} &=& -\frac{L^d}{(4\pi)^{\frac{d+2}{2}}\left( 1-\frac{1}{4}u_0v_0 \right)^{\frac{d+1}{2}}\sqrt{1+\frac{3}{4}u_0v_0}}\left\lbrace  - m^{d+1}\Gamma \left( -\frac{d+1}{2} \right)\sqrt{\left( 1-\frac{1}{4}u_0v_0 \right)\pi} \right. \nonumber \\
&+& \left. a\, m^{d+2}\Gamma \left( -\frac{d+2}{2} \right)+\frac{4m^{\frac{d+2}{2}}\left( 1-\frac{1}{4}u_0v_0 \right)^{\frac{d+2}{4}}}{a^{d/2}}\sum_{n=1}^{\infty}\frac{K_{\frac{d+2}{2}}\left( \frac{2amn}{\sqrt{1-\frac{1}{4}u_0v_0}} \right)}{n^{(d+2)/2}} \right\rbrace .
\end{eqnarray} Note that only the third term in brackets is relevant for the Casimir energy. The other two terms do not contribute because they do not depend on the distance $a$ or because they are linearly related to it. Hence, the Casimir energy for an arbitrary $d$ can be expressed as
\begin{eqnarray}
E^{Cas}_{(iso)}(d) = -\frac{4\, L^d\, a^{-d/2}}{\left( 1-\frac{1}{4}u_0v_0 \right)^{-\frac{d}{4}}\sqrt{1+\frac{3}{4}u_0v_0}}\left( \frac{m}{4\pi} \right)^{\frac{d+2}{2}}\sum_{n=1}^{\infty}\frac{K_{\frac{d+2}{2}}\left( \frac{2amn}{\sqrt{1-\frac{1}{4}u_0v_0}} \right)}{n^{(d+2)/2}}.
\end{eqnarray} Finally, taking $d = 2$ in the expression above, some interesting limits can be analyzed. For example, to the case where  $ma \ll \left( 1-\frac{1}{4}u_0v_0 \right)^{1/2}$  we get
\begin{eqnarray}\label{Ecasisod2lim1}
E_{(iso)}^{Cas} = -\sqrt{\frac{4-u_0v_0}{4+ 3u_0v_0}}\frac{L^2\, \pi^2}{720\, a^3} + \frac{4}{\sqrt{16-8u_0v_0-3u_0^2v_0^2}}\frac{L^2\, m^2}{48\, a} + \cdots ,
\end{eqnarray} where the first term can be recognized as the usual Casimir energy for the massless complex scalar field and the second one as the usual mass term, both fixed by multiplicative Lorentz-violating factors. A quick inspection of the result (\ref{Ecasisod2lim1}) allows us to conclude that the Lorentz-violating coefficient decreases the magnitude of the Casimir energy in the first term. In contrast, the opposite occurs in the second term, where the Casimir energy increase in magnitude. In addition, since the product $u_0v_0$ is supposed to be very small, the Eq. (\ref{Ecasisod2lim1}) can be expanded, up to first order, as
\begin{eqnarray}
    E^{Cas}_{(iso)}=-\frac{\pi ^2 L^2}{720 a^3}+\frac{L^2 m^2}{48 a}+\frac{\pi ^2 L^2 u_0v_0}{1440 a^3}+\frac{L^2 m^2 u_0v_0}{192 a}.
\end{eqnarray} The above expression allow us to see that the net effect of the isotropic Lorentz-violating contribution, expressed by the product $u_0v_0$, is in order to increase the Casimir energy. The behaviour of the Casimir energy for this case is graphically depicted in Fig. (\ref{fig2}). As we can see, when the value of the product $u_0v_0$ increases, the Casimir energy increases as well. Therefore, for too small values of $u_0v_0$, like the current bounds on it ($u_0v_0\approx 10^{-14}GeV$), the curves are practically overlapping.

\begin{figure}[h!]
    \centering
    \includegraphics{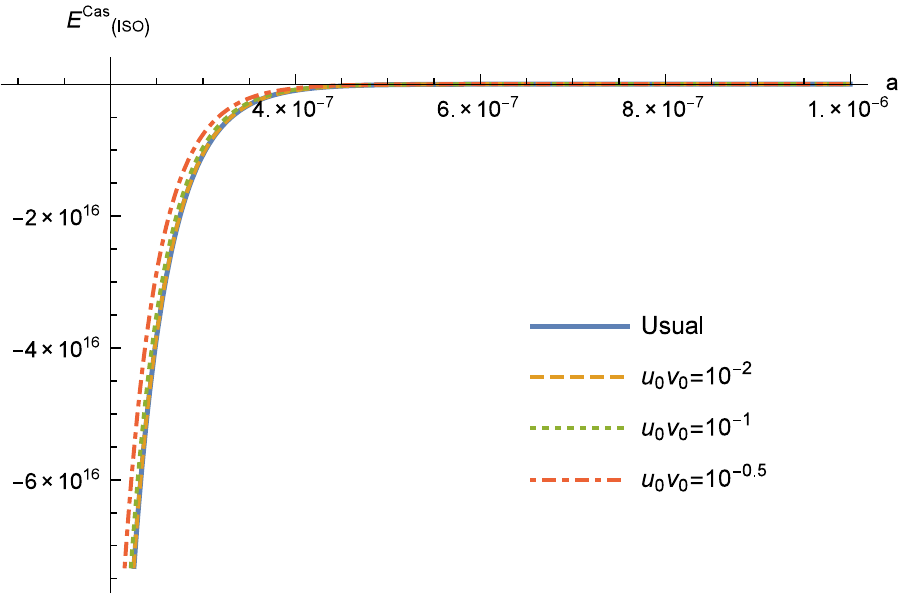}
    \caption{Casimir energy for some values of the product $u_0v_0$. For this plot we have considered $L=1$, $m=10^7$ and $a=(10^{-7},10^{-6})$.}
    \label{fig21}
\end{figure}

On the other extreme, i.e. $ma\gg \left( 1-\frac{1}{4}u_0v_0 \right)^{1/2}$, we have
\begin{eqnarray}\label{ecasisod2lim2}
E^{Cas}_{(iso)} = - \frac{1}{\sqrt{\left( 1+\frac{3}{4}u_0v_0 \right)\sqrt{1-\frac{1}{4}u_0v_0}}} \frac{L^2\, m^2}{8\, \pi^2\, a}\left( \frac{\pi}{m\, a} \right)^{1/2} \exp{\left( -\frac{2\, m\, a}{\sqrt{1-\frac{1}{4}u_0v_0}} \right)}, 
\end{eqnarray} such that the expected exponential decay with respect to the  mass is observed, however shaped by Lorentz-violating factors as well.
Note that the Lorentz-violating coefficient into the exponential factor causes the magnitude of the Casimir energy to decrease faster when compared to the usual case. Accordingly, the Casimir energy must again lead to a small force at the non-relativistic limit. Note that both results \eqref{Ecasisod2lim1} and \eqref{ecasisod2lim2} carry an extra factor of $2$ that takes into account the degrees of freedom of the complex scalar field. Also,  the standard results are recovered when Lorentz violation is ``turned off''. Accordingly, up to first order in $u_0v_0$, Eq. \eqref{ecasisod2lim2} reads
\begin{eqnarray}
    E^{Cas}_{(iso)}=-\frac{L^2 m^2 e^{-2 a m} \sqrt{\frac{1}{a m}}}{8 \pi ^{3/2} a}+\frac{L^2 m^2 e^{-2 a m} \sqrt{\frac{1}{a m}} (4 a m+5)}{128 \pi ^{3/2} a}u_0v_0,
\end{eqnarray} and also here, the Lorentz-violating contribution is strictly positive, thus promoting an increase in the Casimir energy. As we can see from Fig. (\ref{fig21}), the behaviour of the Casimir energy in this regime is very similar to that presented in Fig. (\ref{fig2}).

\section{The anisotropic parity-even sector}\label{sec4}

The anisotropic parity-even sector is in turn parameterized by two pure spacelike
4-vector $U = (0,\mathbf{u})$ and $V = (0,\mathbf{v})$, with $\mathbf{u}\cdot \mathbf{v} = 0$. Thus, the modified Klein-Gordon equation takes the form
\begin{eqnarray}\label{modificadakleingordonparidadepar}
\left[ \Box + m^2+\left( \mathbf{u}\cdot \nabla \right)\left( \mathbf{v}\cdot \nabla \right) \right]\phi = 0.
\end{eqnarray} In order to apply the method of separation of variables for solving Eq. \eqref{modificadakleingordonparidadepar}, two types of solutions will be considered here. The first one will correspond to the case where there is no Lorentz violation in the z-direction, and the second one to the case where only Lorentz violation in the z-direction is taken into account. Let us treat each case separately.

\subsection{No Lorentz violation in the z-direction}

The general situation for this case is reached when $\mathbf{u} = (u_x,u_y,0)$ and $\mathbf{v} = (v_x,v_y,0)$, so that the solution to Eq. \eqref{modificadakleingordonparidadepar} can again be written like Eq. \eqref{solcampoisotrosector}, but now with a new dispersion relation:
\begin{eqnarray}
\omega_{n}(\mathbf{p}) = \sqrt{\mathbf{p}^2+m^2-\left(\mathbf{u}\cdot \mathbf{p}\right)\left(\mathbf{v}\cdot \mathbf{p}\right)+\left(\frac{n\pi}{a}\right)^2},
\end{eqnarray} where $\mathbf{p}=(p_x,p_y)$ as well.

This leads to a new Hamiltonian, that now reads
\begin{eqnarray}\label{hamiltonapenozdir}
H = \frac{1}{2}\sum_{n=1}^{\infty}\int\frac{d^2p}{(2\pi)^2}\left[  a^{\dagger}_n({\mathbf{p}})a_n({\mathbf{p}}) + b^{\dagger}_n({\mathbf{p}})b_n({\mathbf{p}}) + L^22\omega_{n}(\mathbf{p})  \right],
\end{eqnarray} where the annihilation and creation operators $a_n(\mathbf{p})$ and
$a^{\dagger}_n(\mathbf{p})$, respectively, satisfy
\begin{eqnarray}\label{algebraaadaggeranipar}
[a_n(\mathbf{p}),a^{\dagger}_m(\mathbf{q})] = (2\pi)^2\, 2\omega_{n}(\mathbf{p})\, \delta^2(\mathbf{p} - q)\, \delta_{n m}
\end{eqnarray}
with all the other commutation relations involving them being zero and with $b_n({\mathbf{p}})$ and $b^{\dagger}_n({\mathbf{p}})$ obeying a similar algebra.
 
Consequently, the vacuum energy for the massive complex scalar field in the anisotropic parity-even sector with no Lorentz violation in the z-direction is
\begin{eqnarray}
E_{(APE)} = \frac{L^2}{(2\pi)^2}\sum_{n=1}^{\infty}\int d^2p\,  \sqrt{ \mathbf{p}^2 +m^2 -\left(\mathbf{u}\cdot \mathbf{p}\right)\left(\mathbf{v}\cdot \mathbf{p}\right) +\left(\frac{n\pi}{a}\right)^2}.
\end{eqnarray} 

Proceeding as before, we can then write the corresponding Casimir energy for an arbitrary dimension $d$:
\begin{eqnarray}\label{ecasspeddimensional}
E^{Cas}_{(APE)}(d) = -\frac{4\, L^d\, a^{-d/2}}{\left( 1-|\mathbf{u}||\mathbf{v}|\cos{\theta_u}\, \cos{\theta_v} \right )^{d/2}}\left( \frac{m}{4\pi} \right)^{\frac{d+2}{2}}\sum_{n=1}^{\infty}\frac{K_{\frac{d+2}{2}}\left( 2amn \right)}{n^{(d+2)/2}}.
\end{eqnarray} Here $\theta_u$ and $\theta_v$ are respectively the angles that the vectors $\mathbf{u}$ and $\mathbf{v}$ have with respect to the momentum $\mathbf{p}$. 

\begin{figure*}[ht]
\centering
\begin{subfigure}{.5\textwidth}
  \centering
  \includegraphics[scale=0.35]{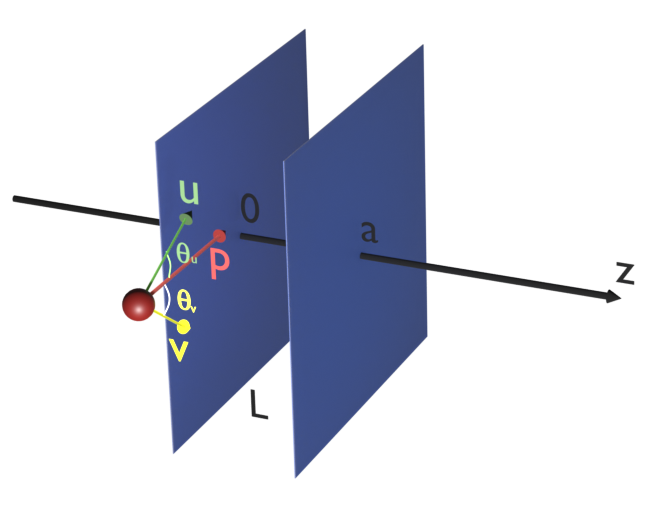}

\end{subfigure}%
\begin{subfigure}{.55\textwidth}
  \centering
  \includegraphics[scale=0.52]{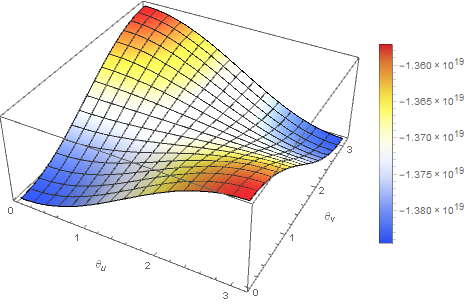}
  
\end{subfigure}
\caption{Casimir energy in terms of the angles between the momentum and the vectors $\mathbf{u}$ and $\mathbf{v}$. For this plot we have considered $L=1$, $m=1$, $|\mathbf{u}|=|\mathbf{v}|=10^{-1}$ and $a=10^{-7}$.}
\label{fig3}
\end{figure*}

For $d=2$, the asymptotic behavior $ma\ll 1$  reads
\begin{eqnarray}\label{ecasapeassintlimit}
E^{Cas}_{(APE)} = \frac{1}{ 1-|\mathbf{u}||\mathbf{v}|\cos{\theta_u}\, \cos{\theta_v}}\left[ -\frac{L^2\pi^2}{720\, a^3}+\frac{L^2m^2}{48\, a} + \cdots \right],
\end{eqnarray} where the terms in brackets can be recognized as the usual Casimir energy and its correction due to the mass. Unlike the isotropic  case, where the usual and mass terms are each corrected by different Lorentz-violating factors, see Eq. \eqref{Ecasisod2lim1}, we see in Eq. \eqref{ecasapeassintlimit} that a single factor breaking the Lorentz symmetry  fixes both terms in the anisotropic parity-even sector. Note that no Lorentz violation effect will be observed if one of the vectors $\mathbf{u}$ and $\mathbf{v}$, or both, are orthogonal to the momentum $\mathbf{p}$ as well. In Fig. (\ref{fig3}), we depict the behaviour of the Casimir energy in terms of the angles between the momentum and the vectors $\mathbf{u}$ and $\mathbf{v}$. As it can be noted, the effect of the Lorentz symmetry violation can be either in order to increase or decrease the Casimir energy. For instance, when $\theta_u=\theta_v=0$ we have a non-vanishing Lorentz contribution and the Casimir energy reaches a minimum value. On the other hand, when $\theta_u=0$ and $\theta_v=\pi$ we have also a non-vanishing Lorentz violating contribution, but the Casimir energy reaches a maximum value instead.

Now, in the limit $ma\gg 1$, although the Eq. \eqref{ecasapeassintlimit} has to be replaced by 
\begin{eqnarray}
E^{Cas}_{(APE)} = -\frac{1}{ 1-|\mathbf{u}||\mathbf{v}|\cos{\theta_u}\, \cos{\theta_v}}\left[ \frac{L^2\, m^2}{8\pi^2 a}\left( \frac{\pi}{m a} \right)^{1/2}e^{-2ma} \right],
\end{eqnarray} we note that some features are preserved. For example, the usual mass exponential decay is fixed by a Lorentz-violating factor that disappear when $\mathbf{u}$ and $\mathbf{v}$ are orthogonal to the momentum $\mathbf{p}$. The general behaviour of the Casimir energy in this particular regime is similar to the one presented in Fig. (\ref{fig3}), being the energy scale the only modification.

\subsection{Lorentz violation only in the z-direction}

In this case, let us assume $\mathbf{u} = u_z\, \mathbf{\hat{z}}$ and $\mathbf{v}= v_z\, \mathbf{\hat{z}}$. As before, the solution to Eq. \eqref{modificadakleingordonparidadepar} is given by Eq. \eqref{solcampoisotrosector} with a new dispersion relation, namely
\begin{eqnarray}\label{reldispapeonlyzdir}
\omega_{n}(\mathbf{p}) = \sqrt{\mathbf{p}^2+m^2+\left( 1-u_zv_z \right)\left(\frac{n\pi}{a}\right)^2}.
\end{eqnarray} Accordingly, the Hamiltonian is once again given by Eq. \eqref{hamiltonapenozdir} with $\omega_{n}(\mathbf{p})$ given by Eq. \eqref{reldispapeonlyzdir}. Therefore, the vacuum energy for the massive complex scalar field in the anisotropic
parity-even sector with Lorentz violation only occurring in the z-direction is
\begin{eqnarray}
E_{(APE)} = \frac{L^2}{(2\pi)^2}\sum_{n=1}^{\infty}\int d^2p\,   \sqrt{\mathbf{p}^2+m^2+\left( 1-u_zv_z \right)\left(\frac{n\pi}{a}\right)^2}.
\end{eqnarray} 

\begin{figure*}[ht]
\centering
\begin{subfigure}{.5\textwidth}
  \centering
  \includegraphics[scale=0.8]{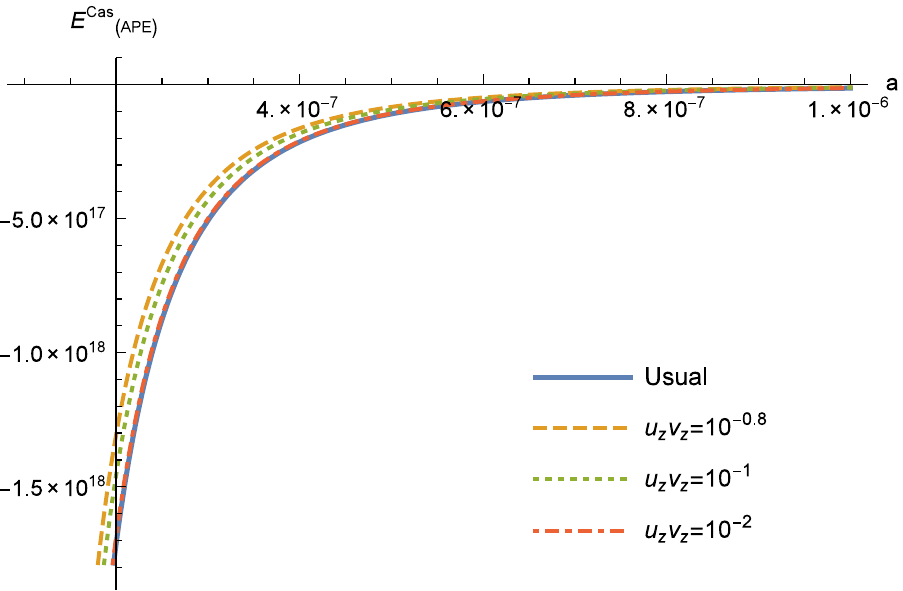}

\end{subfigure}%
\begin{subfigure}{.5\textwidth}
  \centering
  \includegraphics[scale=0.8]{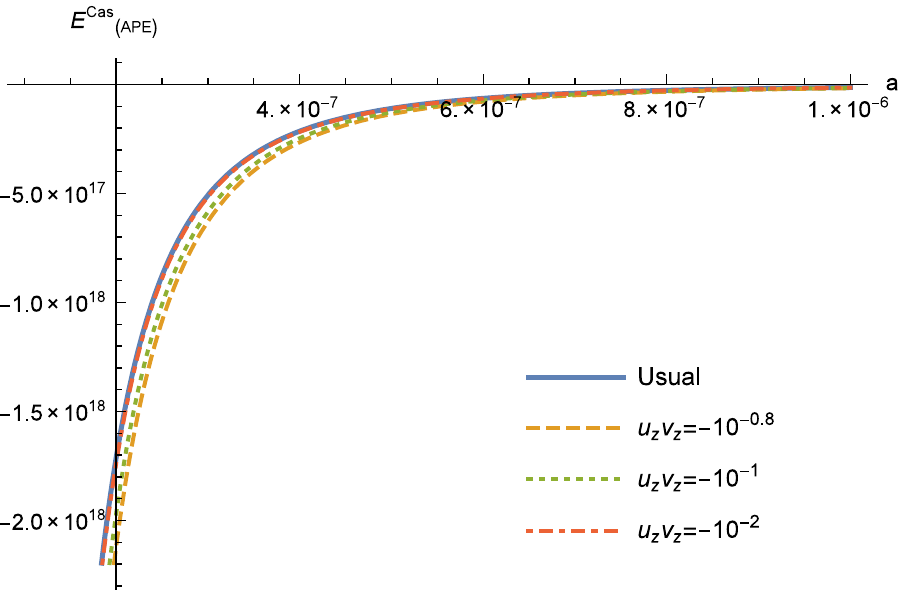}
  
\end{subfigure}
\caption{Casimir energy for parallel and anti-parallel $\mathbf{u}$ and $\mathbf{v}$ vectors. For this plot we have considered $L=1$ and $m=1$.}
\label{fig4}
\end{figure*}

By using dimensional regularization, we can then write
\begin{eqnarray}
E^{Cas}_{(APE)}(d) =  - (1-u_zv_z)^{d/4}\, 4 L^d\, a^{-d/2}\left( \frac{m}{4\pi} \right)^{\frac{d+2}{2}}\sum_{n=1}^{\infty}\frac{K_{\frac{d+2}{2}}\left( \frac{2amn}{\sqrt{1-u_zv_z}}  \right)}{n^{(d+2)/2}}
\end{eqnarray} and by taking $d=2$ we can analyze the limit  $ma \ll \left( 1-u_zv_z \right)^{1/2}$. The result  is
\begin{eqnarray}
E^{Cas}_{(APE)} = \left( 1-u_zv_z \right)^{3/2} \left( -\frac{L^2\pi^2}{720a^3}  \right) + \left( 1-u_zv_z \right)^{1/2} \left( \frac{L^2m^2}{48a}  \right) + \cdots.
\end{eqnarray} 
Similar to the isotropic case, the usual Casimir energy and its mass correction are both fixed by different factors violating the Lorentz symmetry. Also here we can expand the previous expression in terms of $u_zv_z$ up to first order as a way of better understanding the Lorentz violating contribution. Such expansion yields
\begin{equation}
    E^{Cas}_{(APE)}=-\frac{\pi ^2 L^2}{720 a^3}+\frac{L^2 m^2}{48 a}+\frac{\pi ^2 L^2}{480 a^3}u_zv_z-\frac{L^2 m^2}{96 a}u_zv_z.
\end{equation}
From the above expression we can clearly see that the Lorentz violating contributions are not necessarily in order to increase or decrease the Casimir energy, but the increase or decrease depends on some conditions on the parameters of the system. If $m^2<\pi^2/5a$ ($m^2>\pi^2/5a$) the Lorentz violating contribution will increase (decrease) the Casimir energy. Also, for a certain defined range of parameters, the increase or decrease in the Casimir energy may depend on the signal of the product $u_zv_z$. If the vectors $\mathbf{u}$ and $\mathbf{v}$ are parallel, then $u_zv_z>0$. Otherwise, if the vectors $\mathbf{u}$ and $\mathbf{v}$ are anti-parallel, then $u_zv_z<0$. The effects of parallel and anti-parallel $\mathbf{u}$ and $\mathbf{v}$ in the Casimir energy are depicted in Fig. (\ref{fig4}). 

Accordingly, in the limit $ma \gg \left( 1-u_zv_z \right)^{1/2}$, we have
\begin{eqnarray}
E^{Cas}_{(APE)} = - \left( 1-u_zv_z \right)^{3/4}\frac{L^2m^2}{8\pi^2a}\left( \frac{\pi}{ma} \right)^{1/2}\exp \left(-\frac{2ma}{\sqrt{1-u_zv_z}}\right),
\end{eqnarray} or 
\begin{equation}
    E^{Cas}_{(APE)} = -\frac{L^2 m^2 e^{-2 a m} \sqrt{\frac{1}{a m}}}{8 \pi ^{3/2} a} + \frac{L^2 m^3 e^{-2 a m} (4 a m+3)}{32 \pi ^{3/2} (a m)^{3/2}}u_z v_z,
\end{equation} up to first order in $u_z v_z$.
In this regime, the Casimir energy similarly behaves to that one depicted in Fig. (\ref{fig4}), but with a different energy scale.

\section{The anisotropic parity-odd sector}\label{sec5}

In the anisotropic parity-odd sector, the tensor in Eq. \eqref{parametrizacaokemuv} is parameterized by a spacelike and a timelike $4$-vector so that the modified
Klein-Gordon equation take the form
\begin{eqnarray}\label{kleingordonapo}
[\Box + m^2+v_0(\mathbf{u}\cdot\nabla)\partial_0]\phi=0.
\end{eqnarray} Similarly to the previous section, in order to apply the method of separation of variables for solving Eq. \eqref{kleingordonapo}, we will be interested in analyzing the case where there is no Lorentz violation in the z-direction, that is $\mathbf{u}=(u_x,u_y,0)$. The field solution for this situation is again given by Eq. \eqref{solcampoisotrosector} with
\begin{eqnarray}
\omega_{n}(\mathbf{p}) = \sqrt{\frac{v_0^2}{4}(\mathbf{u\cdot \mathbf{p}})^2+\mathbf{p}^2+m^2+\left( \frac{n\pi}{a} \right)^2}.
\end{eqnarray} The Hamiltonian can then be read as
\begin{eqnarray}
H = \sum_{n=1}^{\infty}\int \frac{d^2p}{(2\pi)^2}\frac{1}{4\omega_n(\mathbf{p})^2}\left[ 2\omega_n(\mathbf{p})^2 + v_0\, \omega_n(\mathbf{p})\, (\mathbf{u}\cdot \mathbf{p})\right]\left[  a^{\dagger}_n({\mathbf{p}})a_n({\mathbf{p}}) + b^{\dagger}_n({\mathbf{p}})b_n({\mathbf{p}}) + L^22\omega_n(\mathbf{p})  \right],
\end{eqnarray} where all the annihilation and creation operators satisfy the same algebra as in Eq. \eqref{algebraaadaggeranipar}. Therefore, the vacuum energy for the massive complex scalar field in the anisotropic parity-odd sector with no Lorentz violation effects occurring in the z direction is
\begin{eqnarray}
E_{(APO)} = \frac{L^2}{(2\pi)^2}\sum_{n=1}^{\infty}\int d^2p\left[ \frac{v_0}{2}(\mathbf{u}\cdot \mathbf{p})+ \sqrt{\frac{v_0^2}{4}(\mathbf{u\cdot \mathbf{p}})^2+\mathbf{p}^2+m^2+\left( \frac{n\pi}{a} \right)^2}\,  \right]
\end{eqnarray} By using dimensional regularization and proceeding as before, we can then write the corresponding Casimir energy for an arbitrary
dimension d:
\begin{eqnarray}
E^{Cas}_{(APO)}(d) =  -4\, L^d\, a^{-d/2}\left[ \frac{1}{1+\left( \frac{v_0|\mathbf{u}|\cos \theta_u}{2} \right)^2} \right]^{\frac{d}{2}} \left( \frac{m}{4\pi} \right)^{\frac{d+2}{2}}\sum_{n=1}^{\infty}\frac{K_{\frac{d+2}{2}}\left( 2amn \right)}{n^{(d+2)/2}}.
\end{eqnarray} where $\theta_u$ is again the angle that the vector $\mathbf{u}$ has with the momentum $\mathbf{p}$. For $d=2$, this expression in the limit $ma \ll 1 $ takes the form
\begin{eqnarray}
E^{Cas}_{(APO)} = \frac{1}{1+\left( \frac{v_0|\mathbf{u}|\cos \theta_u}{2} \right)^2}\left[  -\frac{L^2\pi^2}{720\, a^3}+\frac{L^2m^2}{48\, a} + \cdots \right].
\end{eqnarray} 
On the other hand, when $ma\gg \left( 1-\frac{1}{4}u_0v_0 \right)^{1/2}$, we get
\begin{eqnarray}
E^{Cas}_{(APO)} = -\frac{L^2m^2}{8\pi^2a}\left( \frac{\pi}{ma} \right)^{1/2}\frac{e^{-2ma}}{1+\left( \frac{v_0|\mathbf{u}|\cos \theta_u}{2} \right)^2}
\end{eqnarray}

\begin{figure}
    \centering
    \includegraphics{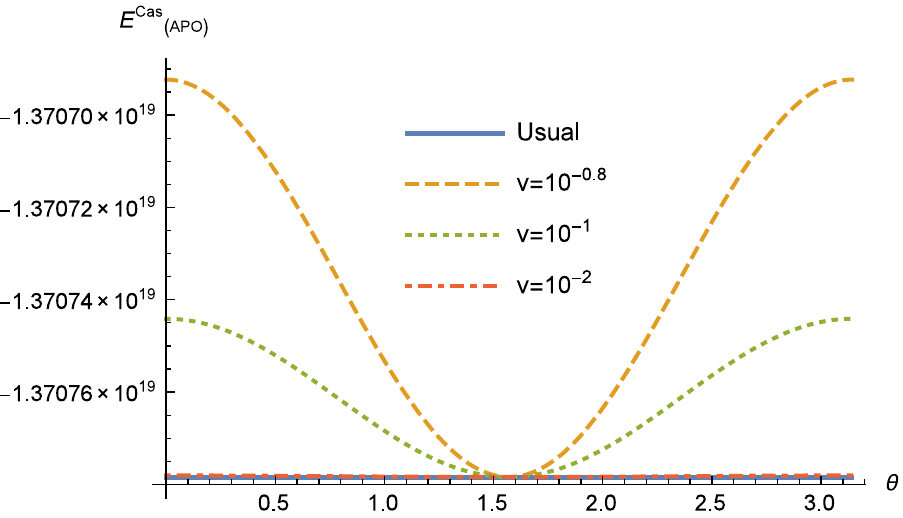}
    \caption{Casimir energy when $ma\ll 1$. For this plot we have considered $L=1$, $m=1$ and $a=10^{-7}$.}
    \label{fig5}
\end{figure}

As we can see, similarly to the anisotropic parity-even case with no Lorentz violation in the z-direction, the anisotropic parity-odd  Casimir energy is just the  usual result for the massive complex scalar field fixed by a Lorentz-violating factor depending on the angle between the vectors $\mathbf{u}$ and $\mathbf{p}$. Furthermore, no Lorentz violation effect will be observed if those vectors are orthogonal to each other. The behaviour of the Casimir energy for the regime $ma\ll 1$ is depicted in Fig. (\ref{fig5}), and an entirely similar behaviour is observed when $ma\gg 1$, being the energy scale the only difference.

%

\section{Conclusion}\label{conclusion}
 
In this paper we investigate the influence of the Lorentz-violating CPT-even contributions of the scalar electrodynamics proposed by Kostelecky and Edwards in the Casimir energy. We have considered the complex scalar field satisfying Dirichlet boundary conditions between two parallel plates separated by a small distance. An appropriate tensor parametrization allwed us to study the Casimir effect in three configurations: isotropic, anisotropic parity-odd and anisotropic parity-even. In each case we have employed dimensional regularization in order to regularize the Casimir energy.
 
For all three configurations we have showed that the Lorentz violating contributions affect the Casimir energy by promoting an increase or decrease on it, depending on the configuration of the system itself. Such an increase or decrease are indeed very small since the current bounds on the Lorentz violating tensor $\hat{K}_c^{\mu \nu}$ are no greater than $10^{-14}GeV$. The increase (decrease) in the Casimir energy consequently yields to a decrease (increase) in the Casimir force. The Casimir force from the Lorentz violating contributions must be at least fourteen magnitude order smaller than the usual Casimir force.

For the isotropic configuration we have found a general solution which was studied in two regimes, namely $ma \ll \left( 1-\frac{1}{4}u_0v_0 \right)^{1/2}$ and $ma \gg \left( 1-\frac{1}{4}u_0v_0 \right)^{1/2}$. Both regimes exhibit a similar behaviour, i.e., the influence of the Lorentz symmetry violation is in order to promote an increase in the Casimir energy, being the energy scale the only difference between the two regimes. Analogously, for the anisotropic parity-odd configuration we have a similar behaviour of the Casimir energy in the regimes where $ma\ll 1$ and $ma\gg 1$. In both cases the Lorentz-violating terms contribute by increasing the Casimir energy, but the amount by which the Casimir is increased depends on the projection of the momentum on the Lorentz-violating vectors.
 
Finally, in the parity-even case, the Lorentz symmetry violation can increase or decrease the Casimir energy. And in this case as well, the amount by which the Casimir energy is increased or decreased depends on the projection of the momentum on the Lorentz-violating vectors. Such a direction dependency on the Casimir energy can be used as a signature to detect small violation in the Lorentz symmetry. It is important to highlight that our results are a direct and important generalization of previous results found in the literature in regard the influence of Lorentz symmetry violation in the Casimir effect.


\section*{Acknowledgments}
\hspace{0.5cm} The authors thank the Funda\c{c}\~{a}o Cearense de Apoio ao Desenvolvimento Cient\'{i}fico e Tecnol\'{o}gico (FUNCAP), Grant no. PNE0112-00085.01.00/16 (JF), and the Conselho Nacional de Desenvolvimento Cient\'{i}fico e Tecnol\'{o}gico (CNPq), Grant no. 200879/2022-7 (RVM) for financial support. R. V. Maluf acknowledges the Departament de F\'{i}sica Te\`{o}rica de la Universitat de Val\`{e}ncia for the kind hospitality.


\end{document}